\documentstyle{article}

\newcommand{\bea}{\begin{eqnarray}}
\newcommand{\eea}{\end{eqnarray}}

\begin{document}

\title{\bf A Rigorous Proof of the Inflationary Spectrum}
\author{Jai-chan Hwang \\
        Department of Astronomy and Atmospheric Sciences \\
        Kyungpook National University, Taegu, Korea}
\maketitle

\begin{abstract}

Fundamental formulae representing the density perturbation spectrums generated
in inflationary scenarios are rigorously proved.
Quantum fluctuations as initial conditions for structure generation
in some inflationary era can be calculated in general scale.
Thus, we could avoid the matching of the large and the small scale results
at the horizon crossing epoch.
This is possible because the perturbed scalar field equation in a
special gauge has some remarkable nice properties.
In the uniform-curvature gauge the contributions from the metric fluctuations
effectively disappear in certain expansion stages, and also we can derive
a simple general solution in the large scale limit.
The equation reduces to the one neglecting the
metric fluctuations in two main bases of inflation scenarios,
namely, the exponential and power law type background expansion;
this makes the quantum generation simple.
The equation leads to a simple form of large scale solution valid for general
scalar field potential; this makes the classical evolution simple.
We also show that the same large scale integral form solution remains valid
for a wide class of generalized gravity theories.

\end{abstract}


\section{Introduction}
                                             \label{sec:Introduction}

An accelerated expansion stage (often termed as inflation)
in the early evolution
phase of our universe provides a natural setting with which we can explain
the currently observed large scale structures by some causal
processes occurred during the acceleration stage.
Usually a scalar field is employed in a concrete model building for the
acceleration stage.
There exist many qualitatively successful scenarios based on its variations
\cite{Linde}.
A further attractive point of the early acceleration phase is that the
ever present microscopic quantum fluctuations (say, in the toy scalar field)
can be expanded into macroscopic quantities during the acceleration phase.
This tempts us to identify them with natural seeds which can be developed into
the observed large scale structure in later time.
The accelerated expansion itself also provides a setting of the
seemingly homogeneous and isotropic observable part of the global
background in which the large scale structures are imbeded.
The improved picture provided by introducing the early acceleration
phase is reasonably simple compared with other attempts.

In this paper we address some points concerning the fundamental formulae
which have been widely used in calculating the density (scalar type mode)
fluctuations generated in the early acceleration phase.
We consider a minimally coupled scalar field and some
representative acceleration phases realized by varying the potential
of the scalar field.
As a background we consider a spatially flat, homogeneous, and
isotropic model universe (Friedmann-Lema{\^i}tre-Robertson-Walker; FLRW).
Parts of the formulae are the following:
\bea
   & & {\Delta \mu \over \mu} \Bigg|_{\rm MDE,HC}
       = \Big| - {2 \over 5} {H \Delta \hat \phi \over \dot \phi} \Big|
       \Bigg|_{\rm UCG,SFDE,LS}, \quad
       {\Delta \mu \over \mu} \Bigg |_{\rm RDE,SS,AMP}
       = \Big| - 4 {H \Delta \hat \phi \over \dot \phi} \Big|
       \Bigg|_{\rm UCG,SFDE,LS}.
					     \label{I}
\eea
where $H (t) \equiv \dot a/a$, $a(t)$ is the cosmic scale factor,
$\mu (t)$ and $\phi (t)$ are the background energy density and the scalar
field,
respectively; an overdot denotes a derivative based on the background
proper-time $t$.
One may find the formulae in Eq. (\ref{I}) are similar to those which appeared
in \cite{Bardeen-etal,Guth-Pi}.
Let us explain the notations used in Eq. (\ref{I}): the subindices indicate
UCG (uniform-curvature gauge), LS (large scale), SS (small scale),
HC (horizon crossing time), MDE (matter-dominated era),
RDE (radiation-dominated era), SFDE (scalar field dominated era),
and AMP (amplitude).
The precise meanings of $\Delta \mu (k,t)$ and $\Delta \hat \phi (k,t)$,
and the reason for taking the absolute values and hats will be explained later
in Eqs. (\ref{P-def}) and (\ref{P-def-x}).
We do not specify any gauge condition for $\Delta \mu (k,t)$'s;
we can consider it as the power spectrum of
Newtonian density fluctuations with the reason
to be explained in \S \ref{sec:Classical}; see below Eq. (\ref{delta-CG}).

In an exponential inflation case, one may often find
$\Delta \hat \phi (k,t) = H/(2\pi)$.
We do have this formula as
\bea
   & & \Delta \hat \phi \Big|_{\rm UCG,EXP,LS} = {H \over 2\pi}
       \Bigg|_{\rm EXP},
					       \label{II}
\eea
where EXP indicates that the quantities are evaluated
during the exponential inflation stage supported by the scalar field;
in the case of power law expansion, the corresponding formula is presented
in eq.(\ref{III}).
Eqs. (\ref{I}) and (\ref{II}) will be derived later.

We would like to point out that derivations of Eqs. (\ref{I}) and (\ref{II})
in
\cite{Bardeen-etal,Guth-Pi,Hawking-Starobinsky,Infl-Pert,Mukhanov-1985,Mukhanov-1988,Mukhanov-etal}
were not complete compared with the presentation in this paper.
(We note that the main results of this paper were previously presented
in \cite{H-QFT,H-UCG}.)
{}For correct analyses, a proper choice of the gauge
(or equivalently the gauge invariant combination) is essential.
Concerning Eq. (\ref{I}) the previous derivation
often used the concept of the exponential inflation.
The exponential expansion leads to a degenerate case where
even different gauge calculations give a similar result.
However, our Eq. (\ref{I}) is generally applicable to scalar
field dominated stage with general potential.
In the case of Eq. (\ref{II}), previous analyses either
neglected the metric perturbations by hand (which is the case of
quantum field in curved spacetime ansatz) or considered the metric
perturbations only in the small scale limit where the metric fluctuation is
negligible again.
(In hindsight we notice that in the exponential expansion case we can manage
the similar result even in other gauges because the case is degenerate.)
In most of the gauge choices the full equation does approach (degenerate to)
Eq. (\ref{no-metric}) in the small scale limit.
However, in that way, the large scale conditions in
Eqs. (\ref{I}) and (\ref{II})
are violated; the numerical justification of the case is
another point \cite{Salopek-etal}.
Notice that Eq. (\ref{I}) is valid when $\Delta \hat \phi (k, t)$ is
evaluated in the large scale where the spatial gradient term is negligible.
In Eq. (\ref{II}) we do evaluate the quantity in the large scale.
In this way, we avoid ad hoc matching of the solutions of the
large and small scale limits at the horizon crossing epoch during the
acceleration phase.
In \cite{H-MSF} we showed that our uniform-curvature gauge analysis is more
straightforward than the zero-shear gauge analyses used in
\cite{Mukhanov-1985,Mukhanov-1988,Mukhanov-etal}.
The derivation of quantum fluctuations in the zero-shear gauge
is difficult and is currently unknown in the non-exponential case.
Later we will show that, in the uniform-curvature gauge,
because of a cancellation
even in the case of power law expansion we can derive corresponding quantum
fluctuations in an exact form; see Eq. (\ref{III}).

Combining Eqs. (\ref{I}) and (\ref{II}) we have
\bea
   & & {\Delta \mu \over \mu} \Bigg|_{\rm MDE,HC}
       = \Big| - {1 \over 5\pi} {H^2 \over \dot \phi} \Big|
       \Bigg|_{\rm EXP}.
					     \label{sp-EXP}
\eea
(The case for radiation dominated era is similar.)
In Eq. (\ref{sp-EXP}) the gauge conditions disappear;
though we have considered the full perturbed metric contributions
in the derivation.
This must be true in any complete analysis.
In other words, the complete analysis using any other gauge
ought to lead to the same result as in Eq. (\ref{sp-EXP});
but not Eqs. (\ref{I}) and (\ref{II}) separately
(separately, they are valid only in the gauge choice mentioned).

In the relativistic perturbation analyses, choosing a proper gauge condition
is very important.
In the analyses involving the scalar field the uniform-curvature gauge
is remarkably convenient;
the uniform-curvature gauge was first introduced in \cite{PRW}.
In the later evolution stage involving the classical ideal fluid,
previously we found
that using some combinations of other gauge conditions is essential for
proper analyses; for example, the perturbed potential and perturbed velocity
in the zero-shear gauge, and perturbed density in the comoving gauge
are closely related with the corresponding Newtonian variables,
and have correct Newtonian behavior in the Newtonian limit \cite{H-MDE,H-IF}.
Each of gauge condition we mention completely fixes the gauge mode.
Thus, any variable under such gauge conditions uniquely corresponds
to a gauge invariant combination.
Without losing generality we can regard the variables under such gauge
conditions as gauge invariant ones.

As a unit, we set $c \equiv 1$.

\subsection{FLRW Background}
                                               \label{sec:BG}

The dynamics of a flat FLRW spacetime is governed by:
\bea
   & & H^2 = {8 \pi G \over 3} \mu, \quad
       \dot H = - 4 \pi G \left( \mu + p \right),
					    \label{BG1}
\eea
where $\mu$ and $p$ are the energy density and the pressure, respectively.
The minimally couled scalar field, $\phi$,
contributes to the fluid quantities as:
\bea
   & & \mu = {1\over 2} \dot \phi^2 + V, \quad
       p = {1\over 2} \dot \phi^2 - V,
					    \label{BG2}
\eea
where $V(\phi)$ is the potential of the scalar field.
{}From Eqs. (\ref{BG1}) and (\ref{BG2}) we can derive
\bea
   & & \ddot \phi + 3 H \dot \phi + V_{,\phi} = 0,
					    \label{EOM}
\eea
where $V_{,\phi} \equiv \partial V/(\partial \phi)$.

{}For an exact de Sitter space, $a \equiv a_0 e^{H(t - t_0)}$ with
$H = {\rm constant}$, supported by the scalar field,
Eqs. (\ref{BG1}) and (\ref{BG2}) imply $\dot \phi = 0 = V_{,\phi}$.
However, as long as the background dynamics is not much affected,
we may allow $V(\phi)$ with nonvanishing $\dot \phi$ while the background
evolution, $H$, is still approximated as de Sitter space.
We consider this as the case for exponential expansion stage.

{}For $a \propto t^{2/(3 + 3 {\rm w})}$ with
${\rm w} \equiv p/\mu = {\rm constant}$
supported by the dynamics of the background scalar field, we have:
\bea
   & & \phi = \sqrt{ 1 \over 6 \pi G ( 1+ {\rm w} ) } \ln{t}, \quad
       V = {( 1 - {\rm w} ) \over 12 \pi G (1 + {\rm w} )^2 }
       e^{-\sqrt{ 24 \pi G ( 1 + {\rm w} ) } \phi }.
                                            \label{POW-BG}
\eea
Thus,
\bea
   & & {H \over \dot \phi} = \sqrt{ 8 \pi G \over 3 ({\rm w} + 1) }.
                                             \label{POW-BG2}
\eea

\subsection{Perturbed equation of motion in FLRW}

The perturbed scalar field equation in the FLRW background
without considering the accompanying metric perturbations is
\bea
   & & \delta \ddot \phi + 3 H \delta \dot \phi + \left( - a^{-2} \nabla^{(3)2}
       + V_{,\phi\phi} \right) \delta \phi = 0,
						\label{no-metric}
\eea
where $\delta \phi ({\bf x},t)$ is a perturbed part of the scalar field,
${\bf x}$ is the comoving spatial coordinate, and
$\nabla^{(3)2}$ is the background comoving three-space Laplacian.
If we do not consider fluctuations in the metric, the perturbed equation can be
derived from the one based on quantum field in curved spacetime
\cite{Birrell-Davies}.
In FLRW background, the scalar field equation $\Box \phi + V_{,\phi} = 0$
leads to
\bea
   & & \ddot \phi + 3 H \dot \phi - a^{-2} \nabla^{(3)2} \phi + V_{,\phi} = 0.
                                \label{QFCS-eq}
\eea
where $\phi = \phi ({\bf x},t)$.
In perturbative approach, considering
$\phi({\bf x},t) = \bar \phi (t) + \delta \phi ({\bf x},t)$
we can derive Eq. (\ref{EOM}) for the background and
Eq. (\ref{no-metric}) for the perturbed part.
Notice that unless we introduce a self interacting potential
Eq. (\ref{no-metric}) has the same structure as Eq. (\ref{QFCS-eq}).

\section{Perturbed Scalar Field in the Uniform-curvature Gauge}
						 \label{sec:UCG}

{\it Gauge condition:}
The most general scalar-type perturbation of the FLRW spacetime is
written as
\bea
   & & d s^2
   = - \left( 1 + 2 \alpha \right) d t^2
       - \chi_{,\alpha} d t d x^\alpha
       + a^2 \delta_{\alpha\beta}
       \left( 1 + 2 \varphi \right) d x^\alpha d x^\beta.
   \label{metric-scalar}
\eea
Without losing generality we choose a spatial gauge condition which
completely fixes the spatial gauge mode; see \S 3 of \cite{H-UCG}.
In this paper we ignore the rotation and the gravitational wave modes
which are completely decoupled in the FLRW background.
{}For the scalar field we let
$\phi ({\bf x}, t) = \phi (t) + \delta \phi ({\bf x}, t)$.
The perturbed order quantities $\alpha ({\bf x}, t)$, $\varphi ({\bf x}, t)$,
$\chi ({\bf x}, t)$, and $\delta \phi ({\bf x}, t)$ are spatially
gauge-invariant, but are temporally gauge dependent; letting any one of
these variables equal to zero can be used as the temporal gauge condition.
The uniform-curvature gauge condition imposes
\bea
   & & \varphi ({\bf x}, t) \equiv 0.
   \label{UCG-condition}
\eea
In this gauge the perturbed part of the three-space curvature
(intrinsic curvature) of a chosen spacelike hypersurface vanishes,
thus justifies its name; see Eq. (\ref{metric-scalar}).
In the uniform-curvature gauge, variables $\alpha({\bf x},t)$ and
$\chi ({\bf x},t)$ completely characterize the scalar type metric fluctuations;
they represent the perturbed lapse function (time-time component
of the metric) and the shear variable of the hypersurface, respectively.
Using the gauge transformation properties of the variables in
\S 2.2 of \cite{PRW} we can construct the following gauge invariant
combinations:
\bea
   & & \delta \phi_\varphi \equiv \delta \phi - {\dot \phi \over H} \varphi,
       \quad
       \alpha_\varphi \equiv \alpha - \left( {\varphi \over H} \right)^\cdot,
       \quad
       \chi_\varphi \equiv \chi - {\varphi \over H}.
   \label{delta-phi-varphi}
\eea
In the uniform-curvature gauge we have
$\delta \phi_\varphi = \delta \phi$, etc.
Thus, the variables in the uniform-curvature gauge can be equivalently
considered as the corresponding gauge invariant combinations of the variables
with $\varphi$.
This is true as long as the chosen gauge condition fixes the gauge mode
completely \cite{PRW}.
Thus, in the uniform-curvature gauge, we have
\bea
   & & \delta \phi = \delta \phi_\varphi, \quad
       \alpha = \alpha_\varphi, \quad
       \chi = \chi_\varphi.
\eea

{\it Equation:}
Since the uniform-curvature gauge condition completely removes the gauge-mode
the equation can be managed into a second order differential equation
which contains only the physical modes.
{}For a derivation, we can use Eqs. (88) and (89) of \cite{PRW};
in that paper we set $8 \pi G \equiv 1$ and equations are presented for
a multi-component case, thus change $\delta \chi_{(i)}$ and $\chi_{(i)}$ into
$\delta \phi ({\bf x},t)$ and $\phi(t)$, respectively.
(A straight derivation from the Einstein' equation can be found in
\cite{H-UCG}.)
Equation (88) of \cite{PRW} is ($k^2 \rightarrow - \nabla^{(3)2}$)
\bea
   & & \delta \ddot \phi_\varphi + 3 H \delta \dot \phi_\varphi
       + \left( - a^{-2} \nabla^{(3)2} + V_{,\phi\phi} \right)
       \delta \phi_\varphi
       = \dot \phi \left( \dot \alpha_\varphi
       - a^{-2} \nabla^{(3)2} \chi_\varphi \right)
       - 2 V_{,\phi} \alpha_\varphi.
                                  \label{PRW-eq88}
\eea
Equation (\ref{PRW-eq88}) is a perturbed part of the equation of motion
for the scalar field.
The terms on the right hand side can be considered as contributions from
the perturbed metric; for interpretations see \S 5 of \cite{H-UCG}.
Since these metric perturbations are caused by the presence of the
perturbed scalar field, there should exist relations between these two
perturbations.
The relations are provided by Einstein's equation as follows.
First two of Eq. (89) in \cite{PRW} are ($\varphi \equiv 0$, $K \equiv 0$):
\bea
   - a^{-2} H \nabla^{(3)2} \chi_\varphi
       + \left( 3 H^2 - 4 \pi G \dot \phi^2 \right) \alpha_\varphi
   &=& 4 \pi G \left( \dot \phi \delta \dot \phi_\varphi
       + V_{,\phi} \delta \phi_\varphi \right),
   \label{PRW-eq89-1} \\
   H \alpha_\varphi
   &=& 4 \pi G \dot \phi \delta \phi_\varphi.
                                 \label{PRW-eq89-2}
\eea
Using Eqs. (\ref{PRW-eq89-1}) and (\ref{PRW-eq89-2})
the right hand side of Eq. (\ref{PRW-eq88}) can be replaced by
$\delta \phi_\varphi$, and we have
\bea
   & & \delta \ddot \phi_\varphi + 3 H \delta \dot \phi_\varphi
       + \Bigg[ - a^{-2} \nabla^{(3)2}
       + V_{,\phi\phi} + 2 {\dot H \over H} \left( 3 H - {\dot H \over H}
       + 2 {\ddot \phi \over \dot \phi} \right) \Bigg] \delta \phi_\varphi = 0.
                                            \label{UCG-eq}
\eea
This is a perturbed scalar field equation which is generally
valid in a flat FLRW background dominated by the background scalar field.
Compared with Eq. (\ref{no-metric}), the additional terms in
Eq. (\ref{UCG-eq}) came from the metric perturbations in Eq. (\ref{PRW-eq88}).
Equations in other gauge choices are more complicated;
for a complete presentation see \cite{H-MSF}.

{\it Limiting cases:}
Equation (\ref{UCG-eq}) has the following interesting consequences.
In the small scale, Eq. (\ref{UCG-eq}) approaches
the same limit of Eq. (\ref{no-metric});
i.e., the spatial gradient term dominates over other terms.
This also happens in many other gauge choices.
Thus, WKB approximation solutions agree in either case.
Also, in the limit of $G \rightarrow 0$ we recover Eq. (\ref{no-metric}).
In the exponential expansion case, where $H$ is constant,
Eq. (\ref{UCG-eq}) becomes Eq. (\ref{no-metric}).
(In an {\it exact} de Sitter background case, from the fact that $\phi$ is
constant both in space and time (\S \ref{sec:BG}),
it follows that  $\delta \phi$ is gauge-invariant; see Eq. (19) of \cite{PRW}.
Thus, since the case of uniform-curvature gauge does,
$\delta \phi({\bf x},t)$ equation in any other
gauge choice should reduce to Eq. (\ref{no-metric}) without potential term;
this may be one point why the previous analysis in different gauges
approximately managed the rigorous result in a near de Sitter space.
We can show that even in the power law expansion case,
which is supported by the background scalar field,
a nontrivial cancellation occurs so that Eq. (\ref{UCG-eq})
reduces to Eq. (\ref{no-metric}) without the potential term.
{}For a general potential, from Eqs. (\ref{BG1}), (\ref{BG2}), and (\ref{EOM})
we can show
\bea
   & & V_{,\phi\phi} + 2 {\dot H \over H} \left( 3 H - { \dot H \over H}
       + 2 {\ddot \phi \over \dot \phi} \right)
       = {H \over a^3 \dot \phi} \left[ {a^3 \dot \phi^2 \over H^2}
       \left( {H \over \dot \phi} \right)^\cdot \right]^\cdot.
                                          \label{cancellation}
\eea
Thus, in the power-law case, using Eq. (\ref{POW-BG2}),
Eq. (\ref{cancellation}) vanishes.
The general condition for such a cancellation to occur is
\bea
   & & a^3 \left( \dot \phi / H \right)^\cdot = {\rm constant},
                                        \label{condition}
\eea
which contains the power-law expansion with Eq. (\ref{POW-BG2}) as a subset;
since $\dot \phi$ goes to zero, the exponential expansion is a subset
of the power-law case.

{\it General asymptotic solutions:}
{}For general $V(\phi)$, the large scale asymptotic solution can be
obtained from Eq. (\ref{UCG-eq}).
We can arrange Eq. (\ref{UCG-eq}) into a compact form as
\bea
   & & {H \over a^3 \dot \phi} \left[ {a^3 \dot \phi^2 \over H^2}
       \left( { H \over \dot \phi} \delta \phi_\varphi
       \right)^\cdot \right]^\cdot
       - a^{-2} \nabla^{(3)2} \delta \phi_\varphi = 0.
                                       \label{UCG-eq2}
\eea
Neglecting the spatial gradient term we have the large scale integral form
solution
\bea
   & & \delta \phi_\varphi ({\bf x}, t)
       = {\dot \phi \over H} \left[ - C ({\bf x})
       + D ({\bf x})\int^t_0 {H^2 \over a^3 \dot \phi^2 } dt \right],
				     \label{UCG-LS-sol}
\eea
where $C ({\bf x})$ and $D ({\bf x})$ are the coefficients of the growing
and the decaying modes, respectively.
Notice that $C ({\bf x})$ and $D({\bf x})$
remain constant independent of the changing background equation of state,
i.e., for general $V(\phi)$.
It is generally possible in other gauge choices to derive the corresponding
general large scale asymptotic solution; see Table 1 of \cite{H-MSF}.
However, a similar general large scale solution is not available in
Eq. (\ref{no-metric}), unless, e.g., the potential term vanishes.

The author of \cite{Mukhanov-1988} introduced a gauge invariant
combination $v$ which is
the same as our $a \delta \phi_\varphi$, thus $a \delta \phi$ in the
uniform-curvature gauge.
(Although the author of \cite{Mukhanov-1988} introduced $v$, which is
$a \delta \phi$ in the uniform-curvature gauge,
the advantage of using $v$ as a uniform-curvature gauge variable
was not realized in \cite{Mukhanov-1988,Mukhanov-etal}.)
In terms of $v$ Eq. (\ref{UCG-eq}) can be written as
\bea
   & & v^{\prime\prime} + \left( k^2
       - {z^{\prime\prime} \over z} \right) v = 0, \quad
       z \equiv {a \dot \phi \over H},
   \label{v-eq}
\eea
where a prime denotes the time derivative based on the conformal time $\eta$,
$d\eta = a^{-1} dt$.
In the large scale limit ($z^{\prime\prime}/z \gg k^2$) we have
$v ({\bf x}, \eta) = c_g ({\bf x}) z + c_d ({\bf x}) z
\int^\eta_0 d \eta / z^2$.
By matching $c_g = - C$ and $c_d = D$, it is equivalent to
Eq. (\ref{UCG-LS-sol}).
In the small scale limit ($z^{\prime\prime}/z \ll k^2$)
we have $v = c e^{ik\eta} + d e^{-ik\eta}$, thus
\bea
   & & \delta \phi_\varphi ({\bf k}, \eta)
       = {1\over a} \Big[ c ({\bf k}) e^{ik \eta}
       + d ({\bf k}) e^{-ik \eta} \Big],
   \label{delta-phi-SS-sol}
\eea
where $c({\bf k})$ and $d ({\bf k})$ are constant coefficients.

\section{Quantum Generation Stage}
                                                        \label{sec:Quantum}

In the exponential and the power law expansion stages we showed that
Eq. (\ref{UCG-eq}) becomes Eq. (\ref{no-metric}), thus with
Eq. (\ref{QFCS-eq}) without self-interaction.
That is, in the exponential and the power law expansion backgrounds the
contribution from the metric fluctuations in Eq. (\ref{UCG-eq}) disappears.
Thus, in these expansion stages
we can adapt the previous (quantum field in curved spacetime)
efforts invested in managing Eq. (\ref{QFCS-eq})
into our case of Eq. (\ref{UCG-eq}).

In our perturbative semiclassical approach, we replace
$\delta \phi_\varphi ({\bf x},t)$
with a quantum (Heisenberg representation) operator
$\delta \hat \phi_\varphi ({\bf x},t)$;
$\delta \hat \phi_\varphi \equiv \delta \hat \phi - (\dot \phi/H) \hat
\varphi$.
Considering that we are in a flat background we have a mode expansion as
\bea
   \delta \hat \phi_\varphi ({\bf x}, t)
   &=& \int {d^3 k \over (2 \pi)^{3/2} }
       \Big[ \hat a_{\bf k} \delta \phi_{\varphi {\bf k}} (t)
       e^{i {\bf k} \cdot {\bf x}}
       + \hat a_{\bf k}^\dagger \delta \phi^*_{\varphi {\bf k}} (t)
       e^{-i {\bf k} \cdot {\bf x} } \Big],
   \label{mode-expansion}
\eea
where $\hat a_{\bf k}$ and $\hat a_{\bf k}^\dagger$ satisfy the standard
commutation relation;
$[\hat a_{\bf k}, \hat a_{{\bf k}^\prime}^\dagger]
= \delta^3 ({\bf k} - {\bf k}^\prime )$
and $0$, otherwise.
The corresponding perturbed metric variables are similarly changed into
operators.
Using Eqs. (\ref{PRW-eq89-1}) and (\ref{PRW-eq89-2}) the operator forms of
the metric variables ($\hat \alpha_\varphi$ and $\hat \chi_\varphi$) can be
expressed as linear combinations of $\delta \hat \phi_\varphi$.
Equation (\ref{UCG-eq}) yields
\bea
   & & \delta \ddot \phi_{\varphi {\bf k}}
       + 3 H \delta \dot \phi_{\varphi {\bf k}} + \Bigg[ {k^2 \over a^2}
       + V_{,\phi\phi} + 2 {\dot H \over H}
       \left( 3 H - {\dot H \over H} + 2 {\ddot \phi \over \dot \phi} \right)
       \Bigg] \delta \phi_{\varphi {\bf k}} = 0.
                                            \label{UCG-eq-k}
\eea

In order to have a proper normalization of the quantum fluctuations, we need
the correct equal time commutation relation.
{}For this we may need a Lagrangian formulation of scalar field perturbation
analysis considering the accompanying metric fluctuations.
Such analysis is available in the literature
\cite{Mukhanov-1988,Mukhanov-etal}.
{}From the Lagrangian formulation (presented by
using $v$) in \cite{Mukhanov-1988}
we can derive the corresponding action in terms of $\delta \phi_\varphi$ as
\bea
   & & S = {1\over 2} \int a^3 \sqrt { g^{(3)} } \Bigg\{
       \delta \dot \phi_\varphi^2
       - {1\over a^2} \delta \phi_{\varphi,\alpha}
       \delta \phi_\varphi^{\;\;|\alpha}
       - {H \over a^3 \dot \phi} \left[ {a^3 \dot \phi^2 \over H^2}
       \left( {H \over \dot \phi} \right)^\cdot \right]^\cdot
       \delta \phi_\varphi^2 \Bigg\} d t d^3 x,
\eea
where $g^{(3)}= 1$ in a flat background, and a vertical bar indicates
the covariant derivative based on $g^{(3)}_{\alpha\beta}$ (which is
$\delta_{\alpha\beta}$ in our flat background).
Considering $S = \int {\cal L} dt d^3 x$, the conjugate momenta is derived as
$\delta \pi_\varphi
\equiv \partial {\cal L} / (\partial \delta \dot \phi_\varphi)
= a^3 \delta \dot \phi_\varphi$.
Thus, from the equal time commutation relation
$ [ \delta \hat \phi_\varphi ({\bf x},t),
\delta \hat \pi_\varphi ({\bf x}^\prime, t) ]
= i \delta^3 ( {\bf x} - {\bf x}^\prime )$
we can derive
\bea
   & & [ \delta \hat \phi_\varphi ( {\bf x}, t), \delta \dot {\hat
\phi}_\varphi
       ({\bf x}^\prime, t) ] = i a^{-3} \delta^3 ({\bf x} - {\bf x}^\prime ).
						\label{commutation}
\eea

The power spectrum is defined as
\bea
   & & {\cal P}_{\delta \hat \phi } ( k , t)
       \equiv [ \Delta \hat \phi ( k , t) ]^2
       \equiv {k^3 \over 2 \pi^2} \left| \delta \phi_{\bf k} (t) \right|^2
       = {k^3 \over 2 \pi^2} \int \langle \delta \hat \phi
       ({\bf x} + {\bf r}, t) \delta \hat \phi ({\bf x},t)
       \rangle_{\rm vac} e^{-i {\bf k} \cdot {\bf r}} d^3 r,
					 \label{P-def}
\eea
where $\langle \rangle_{\rm vac} = \langle {\rm vac}||{\rm vac} \rangle$
indicates a vacuum expectation value with
$\hat a_{\bf k}|{\rm vac}\rangle \equiv 0$.

Eq. (\ref{UCG-eq-k}) can be solved in the following
two background evolution stages.
In both cases Eq. (\ref{UCG-eq-k}) reduces to Eq. (\ref{no-metric}) without
the potential term
\bea
   & & \delta \ddot \phi_{\varphi{\bf k}}
       + 3 H \delta \dot \phi_{\varphi{\bf k}}
       + {k^2 \over a^2} \delta \phi_{\varphi{\bf k}} = 0.
   \label{delta-phi-eq-reduced}
\eea

In an {\it exponential expansion} case Eq. (\ref{delta-phi-eq-reduced})
has an exact solution \cite{Bunch-Davies,Regularization,H-QFT}
\bea
   \delta \phi_{\varphi{\bf k}} ( t )
   &=& c_1 (k) {1\over \sqrt{2k} a} \left( 1 +
       {i \over k \eta} \right) e^{i k \eta}
       + c_2 (k) { 1\over \sqrt{2k} a }
       \left( 1 - {i \over k \eta} \right) e^{- i k \eta}.
                                      \label{EXP-sol}
\eea
The coefficients $c_1(k)$ and $c_2(k)$ are arbitrary functions of $k$ which are
normalized as
\bea
   & & | c_2 (k) |^2 - | c_1 (k) |^2 = 1,
                                         \label{normalization}
\eea
according to Eq. (\ref{commutation}).
Thus $c_1(k)$ and $c_2(k)$ are not completely fixed.
The different choices of $c_1(k)$ and $c_2(k)$ can be interpreted as
the different choices for the vacuum state.
An adiabatic vacuum (in de Sitter space it is often called as
Bunch-Davies vacuum \cite{Bunch-Davies}) corresponds to choosing
\bea
   & & c_2 (k) \equiv 1, \quad c_1 (k) \equiv 0.
                                             \label{BD}
\eea
In the large scale, but in a general vacuum, Eq. (\ref{P-def}) becomes
\bea
   & & {\cal P}^{1/2}_{\delta \hat \phi_\varphi} (k, t) = {H \over 2 \pi}
       \Big| c_2 (k) - c_1 (k) \Big|.
				      \label{II-general}
\eea

In a {\it power law expansion} stage,
with ${\rm w} < - {1\over 3}$, we can derive the exact solution of
Eq. (\ref{delta-phi-eq-reduced}) as \cite{Ford-Parker,POW,H-QFT}
[$a \equiv$ $\left( \eta / \eta_0 \right)^{2/(1 + 3 {\rm w} ) } a_0$]
\bea
   & & \delta \phi_{\varphi{\bf k}} (t) = - { \sqrt{\pi \eta} \over 2 a}
       \left[ c_1 (k) H_\nu^{(1)} ( k \eta)
       + c_2 (k) H_\nu^{(2)} ( k \eta) \right], \quad
       \nu \equiv {3 ( {\rm w} - 1) \over 2 ( 3 {\rm w} + 1) } .
                                      \label{POW-sol}
\eea
In the exponential expansion limit we have ${\rm w} = -1$ ($\nu = {3 \over 2}$)
and Eq. (\ref{POW-sol}) approaches Eq. (\ref{EXP-sol}).
The coefficients $c_1(k)$ and $c_2(k)$ are normalized according to
Eq. (\ref{normalization}).
In the large scale, Eq. (\ref{P-def}) becomes
\bea
   & & {\cal P}^{1/2}_{\delta \hat \phi_\varphi} ( k, t )
       = {1 \over \pi^{3/2} a |\eta|}
       \Gamma \left[ {3({\rm w} - 1) \over 2(3{\rm w} + 1)} \right]
       \left( { k | \eta | \over 2} \right)^{3({\rm w} + 1)/(3 {\rm w} + 1)}
       \Big| c_2 ( k) - c_1 (k) \Big|,
					      \label{III}
\eea
Since $a \propto \eta^{1/2 - \nu}$, Eq. (\ref{III}) does not depend on
any epoch and remains constant in the large scale.
In the exponential expansion limit, we have ${\rm w} = -1$, and
Eq. (\ref{III}) approaches Eq. (\ref{II-general}).
An adiabatic vacuum corresponds to choosing
$c_2 (k) \equiv 1$, $c_1 (k) \equiv 0$.

\section{Classical Evolution Stage}
                                                     \label{sec:Classical}

We consider a later stage of perturbed FLRW model filled with an ideal fluid.
The evolution equation for density fluctuations in the comoving gauge
is well known.
{}From Eq. (41) of \cite{Hwang-Hyun} we have
\bea
   & & \ddot \delta_\Psi + \left( 2 - 6 {\rm w} + 3 c_s^2 \right)
       H \dot \delta_\Psi
       - \left[ {c_s^2 \over a^2} \nabla^{(3)2} + 4 \pi G \mu
       \left( 1 - 6 c_s^2 + 8 {\rm w} - 3 {\rm w}^2 \right) \right]
       \delta_\Psi = 0,
                          \label{delta-CG}
\eea
where $\delta \equiv \delta \mu / \mu$, ${\rm w} \equiv p/\mu$, and
$c_s^2 \equiv \dot p / \dot \mu$; we neglect the imperfect fluid contributions
and assume $K = 0 = \Lambda$.
The comoving gauge fixes the temporal gauge mode completely.
$\delta_\Psi \equiv \delta - 3 H \Psi$ is a gauge invariant combination,
and becomes $\delta$ in the comoving gauge which sets $\Psi \equiv 0$.
$\Psi$ is a velocity related variables, see Eqs. (8) and (44) of \cite{PRW}.
In the Newtonian limit ($p \rightarrow 0$) Eq. (\ref{delta-CG}) reduces
to the corresponding Newtonian perturbation equation.
With this as one reason we suggested that $\delta$ in the comoving gauge
most closely resembles the Newtonian mass density fluctuation
($\delta \varrho/\varrho$) \cite{H-MDE}.

In the large scale we can derive a general integral form solution
[see Eq. (47) of \cite{Hwang-Hyun}]
\bea
   \delta_\Psi ({\bf x},t)
   &=& - {2 \over 3} \nabla^{(3)2} \Bigg[
       {1 \over \dot a^2} C({\bf x})
       \left( 1 - {H \over a} \int^t_0 a dt \right)
       + {1 \over a^2 \dot a} d ({\bf x}) \Bigg],
   \label{LS-sol-CG}
\eea
where $C({\bf x})$ and $d({\bf x})$
are coefficients for the growing and the decaying modes, respectively.
This solution is valid in considering the general time varying $p = p (\mu)$.
We matched the solutions so that $C({\bf x})$ is the same as the one in
Eq. (\ref{UCG-LS-sol});
all solutions are linearly interconnected with each other.
The complete large scale solutions in various gauge are presented in Table 8 of
\cite{H-IF} for an ideal fluid, and in Table 1 of \cite{H-MSF} for a scalar
field.
{}From Table 1 of \cite{H-MSF} we find that $D({\bf x})$ in
Eq. (\ref{UCG-LS-sol}) is $k^2/(aH)^2$ order higher than $d ({\bf x})$.
This means that the dominating decaying mode for $\delta \phi$ in the
uniform-curvature gauge vanishes.

In Newtonian gravity the Poisson's equation relates the potential with the
density field.
In relativistic perturbation analyses
a perturbed potential (curvature) variable, $\varphi$, in the zero-shear gauge
is related to the perturbed density variable in the comoving gauge
through Poisson-like equation; see \S 3.8 of \cite{Hwang-Hyun}.
(The zero-shear gauge condition also fixes the temporal gauge mode completely.
$\varphi_\chi \equiv \varphi - H \chi$ is the corresponding gauge invariant
combination, and becomes $\varphi$ in the zero-shear gauge
which fixes $\chi \equiv 0$.)
Thus, from $- a^{-2} \nabla^{(3)2} \varphi_\chi
= 4 \pi G \delta \mu_\Psi$ we have
\bea
   & & \varphi_\chi ({\bf x},t) = C ({\bf x})
       \left( 1 - {H \over a} \int^t_0 a dt \right) + {H \over a} d ({\bf x}).
                              \label{C-ZSG}
\eea

Physically, we can understand this connection between $\delta \phi_\varphi$
in the early universe and $\delta_\Psi$ in later era as follows.
During the scalar field dominated era, when the scale we are considering
was microscopic, quantum fluctuations in the field ($\delta \hat \phi$)
will simultaneously excite quantum fluctuations in the metric field
(e.g., $\hat \alpha$, $\hat \chi$, or $\hat \varphi$ in other than the
uniform-curvature gauge).
The uniform-curvature gauge is suitable for treating the scalar field
fluctuation.
Due to the accelerated expansion the scale soon becomes macroscopic,
and both fluctuations may become classical.
{}For a coherent picture, there must be a transition from the scalar field
dominated stage to the fluid dominated stage which is often termed as
reheating.
During such transition stage the scale we are considering was far larger than
horizon and the details of the reheating process will not affect the
fluctuations in such a scale.
While the scale remains in the large scale regime we can consider the
information about fluctuation is coded in the metric fluctuations whose
growing mode is characterized by $C({\bf x})$.
Remember that $C ({\bf x})$ is a constant coefficient for solutions which
include the evolving $p(\mu)$ or $V(\phi)$ in the large scale limit.
During the fluid dominated era the generated fluctuations in the metric
($\varphi$) will simultaneously reside together with corresponding fluctuations
in the fluid quantities.
In this way the fluctuation in $\delta$ in fluid era [Eq. (\ref{LS-sol-CG})] is
related to the fluctuations in the scalar field in the early scalar field
dominated era [Eq. (\ref{UCG-LS-sol})].

{}For ${\rm w} = {\rm constant}$, we can derive an exact solution of
Eq. (\ref{delta-CG}) which is
valid for general scale; see \S 3.2 of \cite{H-IF}.
In Fourier space, we have
\bea
   & & \delta_{\Psi {\bf k}} (t) = x^{2 - \beta} \left[ \hat a j_\beta (y)
       + \hat b n_\beta (y) \right], \quad
       y \equiv \sqrt{{\rm w}} x, \quad x \equiv \beta {k \over \dot a}, \quad
       \beta \equiv {2 \over 1 + 3 {\rm w} },
                             \label{gen-sol}
\eea
where $\hat a ({\bf k})$ and $\hat b ({\bf k})$ are constant coefficients.
In the radiation dominated era, since ${\rm w} = {1\over 3}$,
Eq. (\ref{gen-sol}) becomes
\bea
   & & \delta_{\Psi{\bf k}} (t)
       = \sqrt{3} \left[ \hat a \left( {\sin{y} \over y}
       - \cos{y} \right) - \hat b \left( {\cos{y} \over y} + \sin{y} \right)
       \right].
   \nonumber \\
                             \label{RDE-sol}
\eea
In this case Eq. (\ref{LS-sol-CG}) becomes
\bea
   & & \delta_\Psi ({\bf x}, t)
       = - {2 ( 1 + {\rm w}) \over 5 + 3 {\rm w} } {1\over \dot a^2}
       \nabla^{(3)2} C - {2 \over 3} {1\over a^2 \dot a} \nabla^{(3)2} d
       \quad \propto \quad
       t^{ {2 \over 3} { 1 + 3 {\rm w} \over 1 + {\rm w} } } C,
       \quad t^{ { 1 - {\rm w} \over 1 + {\rm w} } } d.
\eea

In the matter dominated era, the spatial gradient term in
Eq. (\ref{delta-CG}) is negligible
as long as the scale we are considering is larger than Jeans scale (sound
horizon) which is negligible compared to the visual horizon; for the Newtonian
analyses see \S 3.3 of \cite{Hwang-Hyun}.
Thus, in the matter dominated era the solution in Eq. (\ref{LS-sol-CG})
remains valid in the regime larger than Jeans scale.
If the scale came inside the visual horizon during the matter dominated era
we have
\bea
   & & \delta_{\Psi{\bf k}} (t) = {2 \over 5}
       {k^2 \over \dot a^2} C_{\bf k}
       + {2 \over 3} {k^2 \over a^2 \dot a} d_{\bf k}
       \quad \propto \quad t^{2 \over 3} C_{\bf k}, \quad t^{-1} d_{\bf k}.
                              \label{MDE-sol}
\eea
Thus, at the (second) horizon crossing epoch in the matter dominated era,
thus $k/(aH) |_{\rm HC}$ $= 1$,
neglecting the decaying mode, we have
\bea
   & & \delta_\Psi |_{\rm HC} = {2 \over 5} C.
                          \label{MDE-delta}
\eea

In the radiation dominated era the spatial gradient term in
Eq. (\ref{delta-CG}) cannot be
neglected near horizon crossing; in the radiation dominated era
the sound horizon scale is
comparable to the visual horizon with $c_s = 1/\sqrt{3}$.
Thus, solution in Eq. (\ref{LS-sol-CG}) is not valid near and after the
horizon crossing due to the violation of the large scale condition.
In the radiation dominated era density fluctuations inside horizon start to
oscillate; see Eq. (\ref{RDE-sol}).
However, its amplitude remains constant.
Considering only the growing mode, from
Eqs. (\ref{LS-sol-CG}) and (\ref{RDE-sol})
we can match the coefficients for the growing modes.
(The same results are obtained by using continuous variables
in sudden jump approximation of transitions between different
background equation of states \cite{Hwang-Vishniac}.)
Thus, in the small scale limit (SS) in the radiation dominated era
the amplitude of $\delta$ becomes
\bea
   & & \delta_\Psi |_{\rm SS, AMP} = 4 C.
                          \label{RDE-delta}
\eea

In this section we reviewed some facts necessary for our proof of
the inflation generated density spectrums which will be presented
in next section.
Concerning the classical evolution stage one can find another way of
presentation in \S IV of \cite{H-QFT}.

\section{Inflationary Spectrum}
                                          \label{sec:Inflation}

{\it An ansatz:}
Analogous to our definition of power spectrum in the quantum context
[Eq. (\ref{P-def})], we can similarly introduce the classical (notice no hat)
power spectrum based on a classical fluctuating field, $f({\bf x},t)$, as
\bea
   & & {\cal P}_{f} ( k, t) \equiv [ \Delta f (k, t) ]^2
       \equiv {k^3 \over 2 \pi^2} \left| f_{\bf k} (t) \right|^2
       = {k^3 \over 2 \pi^2} \int \langle f ({\bf x} + {\bf r}, t)
       f ({\bf x}, t) \rangle_{\bf x} e^{- i {\bf k} \cdot {\bf r} } d^3 r,
					   \label{P-def-x}
\eea
where $f_{\bf k} (t)$ is a Fourier transform of $f ({\bf x},t)$;
$\langle \rangle_{\bf x}$ is a spatial averaging.
This explains why we need the absolute value in Eq. (\ref{I}).
An {\it ansatz} commonly adopted in the literature is to identify
\bea
   & & {\cal P}_{\delta \phi} (k, t) \equiv {\cal P}_{\delta \hat \phi} (k, t).
                                      \label{ansatz}
\eea
In order to distinguish this ansatz we keep hats in
Eqs. (\ref{I}) and (\ref{II}).
The power spectrum of the density fluctuations in Eq. (\ref{P-def-x})
is closely related to the mass fluctuations; see Eq. (69) of \cite{H-QFT}.
In this context, ${\cal P}_\delta (k,t)$ is often used in characterizing
density fluctuations.

{\it Generated spectrum:}
The power spectrums, $\Delta \hat \phi_\varphi (k, t)$ for the exponential
inflation and the power law inflation are derived in
Eqs. (\ref{II-general}) and (\ref{III}), respectively.
Both expressions remain constant in time.
By using an adiabatic vacuum, Eq. (\ref{II-general}) becomes
Eq. (\ref{II}) for an exponential expansion.

Although they are evaluated in different eras and different gauges,
the same $C ({\bf x})$ appears in Eqs. (\ref{UCG-LS-sol}), (\ref{MDE-delta}),
and (\ref{RDE-delta}).
This reflects the fact that through linear evolution we do not have
structure formation; information about the spatial structure is coded in
$C({\bf x})$ and the coefficient for the decaying mode.
Combine Eqs. (\ref{UCG-LS-sol}), (\ref{MDE-delta}), and (\ref{RDE-delta}).
Replace $\delta_{\Psi{\bf k}} (t)$ and $\delta \phi_{\varphi{\bf k}} (t)$
with the quantities characterizing the fluctuations in
Eqs. (\ref{P-def}) and (\ref{P-def-x}).
Then, using an ansatz in Eq. (\ref{ansatz}), we can derive Eq. (\ref{I}).
This completes our argument presented in the introduction.

Equation (\ref{sp-EXP}) is based on a widely accepted vacuum choice;
see below Eq. (\ref{normalization}).
In fact, we can write it in an arbitrary vacuum.
The corresponding density spectrum for the exponential and the power law
inflation can be found from Eqs. (\ref{I}), (\ref{II-general}), and
(\ref{III}):
\bea
   {\Delta \mu \over \mu} \Bigg|_{\rm MDE,HC}
   &=& \Big| - {1 \over 5\pi} {H^2 \over \dot \phi} \Big|
       \times \Big| c_2 (k) - c_1 (k) \Big| \Bigg|_{\rm EXP},
   \label{gen-sp-EXP} \\
   {\Delta \mu \over \mu} \Bigg|_{\rm MDE,HC}
   &=& \Big| - \sqrt{ 8 \pi G \over 3 ( {\rm w} + 1) }
       {H (3 {\rm w} + 1) \over 5 \pi^{3/2} }
       \Gamma \left[ {3 ( {\rm w} - 1)
       \over 2 ( 3 {\rm w} + 1 ) } \right]
       \left( {k |\eta| \over 2} \right)^{3({\rm w} + 1)/(3 {\rm w} + 1)} \Big|
       \Big| c_2 (k) - c_1 (k) \Big| \Bigg|_{\rm POW},
   \label{gen-sp-POW}
\eea
where $c_1(k)$ and $c_2(k)$ only subject to a condition in
Eq. (\ref{normalization}).
Corresponding results can be given in the radiation dominated era case.
We used the initial condition which is generated from quantum fluctuations
evaluated in the large scale; Eqs. (\ref{II-general}) and (\ref{III}).
If we put the constraint that in the small scale the solution
should have corresponded to the positive frequency adiabatic vacuum
[see below Eq. (\ref{normalization})],
we recover Eq. (\ref{sp-EXP}) in the exponential case.
It is often accepted that the observable part of the universe, thus the
observed large scale structure in it, may come out from the range where
the adiabatic vacuum choice is relevant \cite{Ford-Parker,Regularization}.

{}From Eqs. (\ref{P-def-x}) and (\ref{MDE-sol}) we have
\bea
   & & {\Delta \mu \over \mu} \Bigg|_{\rm HC}
       = {\dot a^2 \over \sqrt{2} \pi} k^{-1/2}
       \left| \delta_{\bf k} (t) \right|.
\eea
Following the convention
$|\delta_{\bf k} (t) |^2 \equiv A (t) k^n$,
Eqs. (\ref{gen-sp-EXP}) and (\ref{gen-sp-POW})
in the adiabatic vacuum situation lead to
\bea
   & & n |_{\rm EXP} = 1, \quad
       n |_{\rm POW} = {9 {\rm w} + 7 \over 3 {\rm w} + 1}.
					      \label{n}
\eea
The first one is known as Zel'dovich spectrum; $\ln{k}$ dependence may still
arise as we evaluate the time varying $H^2/\dot \phi$ (especially $1/\dot
\phi$)
in specific inflation models which may deviate (not much of the expansion
dynamics, but the physical state of $\dot \phi$; see \S \ref{sec:BG})
from the exact de Sitter case; in an exact exponential expansion limit
Eq. (\ref{sp-EXP}) diverges.
The spectrum for the power law case in Eq. (\ref{n}) is the same as
the one in \cite{Lucchin-Matarrese}.

\section{Generalized Gravity Theories}
                                            \label{sec:GGT}

Using the uniform-curvature gauge a class of generalized gravity theories
also yields the simple form of the large scale solution.
The following Lagrangian includes various types of generalized gravity theories
\cite{H-GGT}
\bea
   & & {\cal L} = \sqrt{-g} \left[ {1\over 2} f (\phi, R)
       - {1\over 2} \omega (\phi) \phi^{;a} \phi_{,a} - V(\phi) \right].
                               \label{GGT-Lagrangian}
\eea
It includes $f(R)$ gravity, scalar-tensor theory, nonminimally coupled scalar
field, induced gravity, etc, as subclasses.
The general theory included in this Lagrangian is a two component system.
Thus, in general, we may end up with a fourth order differential equation
which describes the scalar type perturbations.
However, each subclass we mentioned corresponds to one component system.
The
perturbation analysis can be made in a unified manner similarly as in the case
of minimally coupled scalar field \cite{H-GGT,PRW,Mukhanov-etal}.
Part of the reason why we have such a simple treatment is that there exists
a conformal transformation which transforms each case of
generalized gravity theory into Einstein's gravity supplemented by
a minimally coupled scalar field with a special potential.
{}For a perturbation treatment, however, we can proceed it without using
the conformal transformation.
The analyses were previously made
based on the zero-shear gauge \cite{H-GGT,PRW}.
In the following we present a translation of the solutions derived in the
zero-shear gauge into ones in the uniform-curvature gauge \cite{Hwang-Minn}.

In the zero-shear gauge we derived the large scale integral form solutions
which are valid
for general $V(\phi)$ [see Eqs. (85) and (86) of \cite{PRW}]:
\bea
   & &
       {\delta \phi_\chi \over \dot \phi}
       = {\delta F_\chi \over \dot F}
       = - C {1 \over a F} \int^t_0 a F dt + {1\over aF} d, \quad
       \varphi_\chi = C \left( 1 - {H \over aF} \int^t_0 a F dt \right)
       + {H \over aF} d,
                            \label{ZSG-sol}
\eea
where $F \equiv f_{,R}$.
The coefficients $C({\bf x})$ and $d({\bf x})$ are matched to the solutions
in the minimally coupled
scalar field and the ideal fluid; see Eqs. (\ref{UCG-LS-sol}) and
(\ref{C-ZSG}).
$\delta \phi$ in the uniform-curvature gauge is the same as
$\delta \phi_\varphi$ in Eq. (\ref{delta-phi-varphi}).
Evaluating $\delta \phi_\varphi$ in the zero-shear gauge and using
the solutions in Eq. (\ref{ZSG-sol}) we can derive the large scale
solutions in the uniform-curvature gauge [we have
$\delta F_\varphi \equiv \delta F - (\dot F/H) \varphi$]:
\bea
   & & \delta \phi_\varphi = - {\dot \phi \over H} C({\bf x}), \quad
       \delta F_\varphi = - {\dot F \over H} C({\bf x}).
                             \label{UCG-GGT-sol}
\eea
Disappearence of the decaying mode in the uniform-curvature gauge
means that the dominating decaying mode vanishes.
The decaying mode in Eq. (\ref{UCG-LS-sol}) is higher order
(in large scale expansion) compared with the one in other gauge;
see below Eq. (\ref{delta-CG}) and \cite{H-MSF}.
$\delta \phi_\varphi$ in Eq. (\ref{UCG-GGT-sol}) is simpler than
$\delta \phi_\chi$ in Eq. ({\ref{ZSG-sol}).
Neglecting the decaying mode, even in the generalized gravity theories we have
\bea
   & & C ({\bf x}) = - {H \over \dot \phi} \delta \phi_\varphi ({\bf x},t),
   \label{C-delta-phi}
\eea
which is free of $F$ and $\omega$, thus, is the same as Eq. (\ref{UCG-LS-sol})
valid for the minimally coupled scalar field.

The solutions in the uniform-curvature gauge above were obtained by using
the known solutions in the zero-shear gauge.
Using the complete set of equations presented in \S 4.2 of \cite{PRW} we can
directly derive the equation for $\delta \phi_\varphi$ and its asymptotic
solutions in the generalized $f(\phi, R)$ gravity;
recently, we present a thorough analyses in \cite{H-GGT-UCG}.
Whether the uniform-curvature gauge could allow the analytic derivation
of quantum fluctuations in the
context of nonminimally coupled scalar field or other generalized gravity
theories is an interesting question.
We would like to address this subject in a future occasion.

\section{Summary}
                                            \label{sec:Summary}

We summarize the result in the following.

\begin{itemize}
\item
Perturbative semiclassical approximation simultaneously treats
both the perturbed field and the perturbed metric quantum mechanically.
By appropriate choice of the gauge, the perturbative approach is easily
tractable and becomes more self-consistent than the quantum field
analysis in curved spacetime.

\item
We show the uniform-curvature gauge provides the most convenient analysis
in treating the scalar field and a class of generalized gravity theories.

\item
In the large scale, the equation in the uniform-curvature gauge yields
an integral form solution for a general $V(\phi)$; Eq. (\ref{UCG-LS-sol}).
This is not possible if we neglect the metric perturbations.
Similar large scale integral form solutions are available in the other gauges,
but they are  more complicated; for a thorough analysis in other gauges,
see \cite{H-MSF}.
The general solution in the small scale limit is presented in
Eq. (\ref{delta-phi-SS-sol}).

\item
The perturbed scalar field equation in the uniform-curvature gauge exactly
reduces to the one which neglects the metric perturbations in the
exponential and the power law expanding backgrounds.
In these cases we consider the background dynamics is supported by the
background scalar field.
Meanwhile, in the approach of the quantum field theory in curved spacetime
the background has been assumed to be given by an external source.

\item
The quantum initial conditions can be derived in general scale.
Thus, we can avoid any small and large scale matching at horizon crossing
epoch;
\S \ref{sec:Quantum}.

\item
The previously known and also widely used formulae are derived
in exact settings; Eqs. (\ref{I}), (\ref{sp-EXP}), and (\ref{gen-sp-POW}).
This enables us to evaluate Newtonian density perturbations in terms of
quantum fluctuations generated in an inflation era;
Eqs. (\ref{sp-EXP}), (\ref{gen-sp-EXP}), and (\ref{gen-sp-POW}).
Taking a suitable gauge condition is essential for making the above
simple analysis possible, particularly in quantum generation processes.

\item
Our result in Eq. (\ref{I}) can be generally applied to
inflation models based on the minimally coupled scalar field.
Notice that in Eq. (\ref{I}), $V(\phi)$ is general as long as the
dynamics is governed by the scalar field.

\item
The density spectra in Eq. (\ref{n}), are not only
characterized by the background expansion dynamics at the quantum generation
stage, but also depend on the choice of the natural vacuum state;
Eqs. (\ref{gen-sp-EXP}) and (\ref{gen-sp-POW}).

\item
We used a semiclassical method in quantizing only the perturbed parts.
In a related context Eq. (\ref{ansatz}) remains as an ansatz.
An important related issue of the classicalization of the quantum fluctuations
is not addressed in this paper.

\item
We presented the simple forms of the large scale integral solutions valid
in a class of generalized gravity theories in the uniform-curvature gauge;
Eq. (\ref{UCG-GGT-sol}).
Remarkably, considering the growing mode, the same solution for minimally
coupled scalar field in Eq. (\ref{UCG-LS-sol}) remains valid in a class of
generalized gravity theories in Eq. (\ref{GGT-Lagrangian}).

\item
Equation (\ref{UCG-eq}) is valid for a general scalar field dominated model.
Thus, it will be interesting to apply it to situations where
the terms appended through the metric fluctuations do not vanish.
This is left as a future study.

\end{itemize}

One lesson we notice from this study is that, although the final
physical result should not depend on any gauge choice we are working on, in
managing an actual situation, it is very important to choose
an appropriate gauge which suits the problem.
Indeed, the gauge choice is a freedom we have.
In this regards, the gauge ready method proposed in
\cite{Bardeen-1988,PRW} is convenient because it allows an easy
adaptation of the perturbed system into any gauge choice and also allows
an easy translation between solutions in different gauge choices.
It is known that the density fluctuations in the comoving gauge
($\delta \mu |_{\rm CG}$), the potential (or curvature) and the velocity
fluctuations in the zero-shear gauge
($\varphi |_{\rm ZSG}$ and $\delta v |_{\rm ZSG}$)
most closely resemble the Newtonian counterparts of density, potential, and
velocity fluctuations, respectively; see \cite{H-MDE}.
{}From our study we may identify one additional case: the scalar field
fluctuations in the uniform-curvature gauge ($\delta \phi |_{\rm UCG}$)
closely resemble the field fluctuations based on the pure background metric.
These fluctuations in the mentioned gauges can be regarded as the
gauge-invariant
concepts.
That is we have $\delta \mu |_{\rm CG} = \delta \mu_\Psi$,
$\varphi |_{\rm ZSG} = \varphi_\chi$, $\delta v |_{\rm ZSG} = \delta v_\chi$,
and $\delta \phi |_{\rm UCG} = \delta \phi_\varphi$.

Now, we summarize the large scale structure generation and
the evolution process.
The microscopic quantum fluctuations are expanded into the macroscopic scale
during inflation stage and generate the
large scale (and macroscopic) fluctuations
in the perturbed scalar field.
The scalar field dominated stage can be easily handled
in the uniform-curvature gauge, $\S\ref{sec:Quantum}$.
The simultaneously excited metric fluctuations can be characterized by
curvature (potential) fluctuations in the fabric of the metric in the
zero-shear gauge, Eq. (\ref{C-ZSG}).
This can be transformed into the Newtonian (the comoving gauge one) density
fluctuations later, leading to the results in
Eqs. (\ref{sp-EXP}), (\ref{gen-sp-EXP}), and (\ref{gen-sp-POW})
in some inflation stages we consider.
Most of the time the fluctuation scales stay in the large scale (outside
horizon
for general $p$, and outside Jeans scale for the matter dominated era).
In such scales the evolution of fluctuations is characterized by
a conserved quantity $C ({\bf x})$ appearing in the large scale solutions.
$C$ appears as an integration constant for the growing mode in every variable.
The curvature (potential) variable, $\varphi$, in many gauge choices is
conserved in the large scale and can be identified with $C$.
[$\varphi$ is $C$ on
scales larger than horizon in most of the gauge choices; the exceptions are
the zero-shear gauge where we have Eq. (\ref{C-ZSG}) and the uniform-curvature
gauge where we take $\varphi \equiv 0$ as the gauge condition.]
During scalar field dominated inflation era, the spatial information about
$\delta \phi ({\bf x}, t)$ (generated from vacuum quantum fluctuations) in the
uniform-curvature gauge is encoded into $C ({\bf x})$, Eq. (\ref{UCG-LS-sol}).
Ignoring the transient mode we have
\bea
   & & C ({\bf x}) = - {H \over \dot \phi} \delta \phi_\varphi ({\bf x},t),
\eea
which remains valid even in the generalized gravity theories
in Eq. (\ref{C-delta-phi}).
$C$ carries the spatial information about the initial condition.
Later the same $C$ can be translated (decoded) into the (growing mode part)
information about $\varphi$ in the zero-shear gauge [Eq. (\ref{C-ZSG})] or
$\delta$ in the comoving gauge [Eq. (\ref{LS-sol-CG})] and in fact any variable
in any gauge.
In an ideal fluid case, $\delta \mu_\Psi / \mu$, $- \varphi_\chi$,
$v_\chi$, and $\delta T_\Psi / T$ play the roles of the Newtonian
relative density fluctuation ($\delta \varrho / \varrho$),
potential fluctuation ($\delta \Phi$), velocity fluctuation ($\delta v$),
and the relative temperature fluctuations in the cosmic microwave background
radiation ($\delta T/T$), respectively \cite{H-MDE}.
[$v$ is related to $\Psi$ as $v \equiv -(k/a) \Psi/(\mu + p)$.
$\Psi$ is a frame invariant velocity related variable;
see \S 2.1.2 of \cite{PRW}.]
In the matter dominated stage, ignoring the decaying modes,
we have \cite{H-MDE} [see Eqs. (\ref{MDE-sol},\ref{C-ZSG})]:
\bea
   & & {\delta \varrho \over \varrho} = {2 \over 5}
       \left( {k \over a H} \right)^2 C, \quad
       \delta \Phi = - {3 \over 5} C, \quad
       \delta v = - {2 \over 5} \left( {k \over aH} \right) C, \quad
       {\delta T \over T} = {1\over 5} C.
   \label{pert-sols}
\eea
Through the linear evolution the spatial structures are preserved.
$C({\bf x})$ encodes the spatial structure of the growing mode
in the linear regime.
We would like to remark that this is {\it a} way of describing
the structure generation process which is simple and concrete.



\end{document}